# Room Temperature RF Sputtering of Mixed Ionic and Electronic Conductor $Nd_2Ni_{0.8}Cu_{0.2}O_{4+\delta}$ films

N. Coppola[1], M. Paone[1], H.S. Ur Rehman[1], S. Scarnicci[1], G. Carapella[2], A. Guarino[3], M. Tkalcevic[1*], L. Calcagnile[4], G. Quarta[4], A. Galdi[1], L. Maritato[1]

[1]Dipartimento di Ingegneria Industriale, Università degli Studi di Salerno, Via Giovanni Paolo II 132, 84084 Fisciano (SA), Italy

[2]Dipartimento di Fisica "E. R. Caianiello", Università degli Studi di Salerno, via Giovanni Paolo II 132, I-84084 Fisciano (SA), Italy [3] CNR-SPIN, c/o University of Salerno, 84084 Fisciano (Salerno), Italy

[4]CEDAD-Centre of Applied Physics, Dating and Diagnostics, Department of Mathematics and Physics "Ennio de Giorgi", University of Salento and INFN-Lecce Section, Lecce, Italy

* Current affiliation: Institut Ruđer Bošković, Bijenička cesta 54, 10000, Zagreb, Croatia

## Abstract

Lowering the operating temperature of Solid Oxide Fuel Cells (SOFCs) is essential for improving durability and enabling large-scale commercialization. Mixed ionic–electronic conductors (MIECs) of the Ruddlesden–Popper family, such as $Nd_2Ni_{1-x}Cu_xO_{4+\delta}$ (NNCO), offer attractive cathode properties due to their high oxygen transport and favourable defect chemistry. In this work, we investigate the fabrication of $Nd_2Ni_{0.8}Cu_{0.2}O_{4+\delta}$ thin films using a room-temperature RF sputtering process followed by moderate-temperature annealing. To simplify deposition, a single stoichiometric target was employed, despite the compositional challenges posed by elements with different sputtering yields. We examine the effect of sputtering power density on phase formation and film stoichiometry through X-ray diffraction, Rutherford Backscattering Spectrometry, Energy-Dispersive X-ray Spectroscopy, and temperature-dependent resistivity measurements.

Increasing the sputtering power density strongly reduces the presence of spurious phases and promotes stabilization of the desired n=1 Ruddlesden–Popper phase. Films deposited at 230 W (3.1 W·cm$^{-2}$) exhibit a predominant NNCO structure, elemental ratios close to nominal composition, and resistivity values consistent with bulk materials. These findings demonstrate that high-power sputtering combined with ex-situ annealing enables the production of NNCO thin films suitable for SOFC cathodes. The results support the potential of PVD-based approaches for scalable fabrication of advanced SOFC components.

## Introduction

The focus of the research on solid oxide fuel cells has significantly raised during the last years due to the interesting characteristics and performance this technology has shown like fuel flexibility, convertibility and high efficiency[1–4]. In order to ameliorate and let this technology marketable, the scientific community has pointed the attention on the weak points of the process, of them, the biggest and most impactful parameter is the operating temperature, typically from 800 to 1000°C. These temperatures are needed because of the using of non-precious materials as electrodes and to ensure a sufficient ionic conductivity in the electrolyte. However, these extreme conditions lead to a fast degradation of the components reducing the long-term durability. Consequently, a major scientific effort is directed toward lowering the operating temperature to the intermediate range (600-800°C)[5–7].

One of the major bottlenecks of the process is represented by the cathode where the oxygen-reduction reaction (ORR) occurs[8]. At the present, one of the most common used materials is the perovskite lanthanum-strontium manganite (LSMO) which possess good electronic conductivity and poor ionic transport. In the case of LSMO, the electrochemical reaction is restricted to the Triple Phase Boundaries (TPB), the microscopic points where the three species, gas (oxygen), the electrolyte (ionic conductivity), and the electrode (electrons) meet. This limitation severely curtails the number of $O^{-2}$ ions involved in the electrochemical reactions. With the aim to overcome such a restriction, several researches have focused on the possible use as cathodes in SOFCs of a new class of materials, the so-called Mixed Ionic-Electronic Conductors (MIECs)[9–11]. MIEC materials can conduct both oxygen ions and electrons, extending the active ORR sites to the entire cathode's volume and finally resulting in a larger number of $O^{-2}$ ions reaching the electrolyte. Recently, a class of MIEC materials attracting strong research interest has been that of the perovskite-based Ruddlesden-Popper series, with general formula $A_{n+1}B_nO_{3n+1}$, where A is a rare earth and B is a transition metal. In particular, the Ni based compounds in this series with n=1 and A = Nd, the $Nd_2NiO_{4+\delta}$ (NNO), has been considered as a good SOFC cathode candidate primarily because of its peculiar oxygen diffusion mechanism mainly based on an oxygen push-pull process[12], different from the usual oxygen migration via vacancies mechanism found in LSMO. Recent results[13] have demonstrated that the high concentration of interstitial oxygen favours the ionic transport in bulk NNO, while Density Functional Theory (DFT) studies have shown that the partial Ni substitution with Cu in $Nd_2Ni_{1-x}Cu_xO_{4+\delta}$, (NNCO), decreases the oxygen hopping activation energy for x up to 0.2, confirming the observation, in standard SOFC with screen printed NNCO cathode, of important decrease in the overall cell polarization resistance[12,14]. Moreover, the choice of nickelates as cathode materials is corroborated by the fact that they do not contain cobalt or alkali earths that tend to segregate on the electrode's surface and diminish the cell performance[15].

The use of innovative SOFC fabrication process based on Physical Vapour Deposition (PVD) techniques has been lately proposed by several research groups[7,16–18], showing the possibility to improve the cell final electrochemical performances and durability. PVD techniques are generally difficult to scale up to industrial production lines mainly because of the need to control and stabilize, during the deposition process, the high temperature over large substrate area. A successful and innovative process has been recently proposed[17,19–21] in which a room temperature RF-Sputtering deposition is followed by a separate annealing process at moderate temperatures. Industrial size SOFC in which this technique has been used to fabricate the Gd-doped Ceria buffer layer at the electrode/electrolyte interface, have shown improvements up 30% in the output current density at 800 mV and 650 °C, compared to standard industrial cells with screen printed buffer layer.

Following these studies, we decide to investigate the adoption of this technique also to fabricate NNCO thin films in view of their application as cathodes in SOFC. To simplify the process, we choose to use a single sputtering target made of NNCO with $x = 0.2$. When depositing from target presenting elements with different sputtering yields, the composition of the obtained films can substantially vary as a function of the sputtering parameters[22,23]. In this work, we focus our attention over the influence of the sputtering power density on the final stoichiometry of the produced NNCO films. The deposited films are characterized by X-Ray Diffraction (XRD), Energy Dispersive X-Ray Spectroscopy (EDX), Rutherford Back-Scattering (RBS) and electric resistivity vs. temperature ($\rho(T)$) measurements. The obtained results have shown the possibility to produce samples having the predominant Ruddledsen-Popper desired phase, opening the way to future developments in view of the industrial scaling up of the proposed technique for the fabrication of large-scale cathodes in SOFC.

# Experimental Methods

## *Sample preparation*

$Nd_2Ni_{0.8}Cu_{0.2}O_{4+\delta}$ (NNCO) thin films have been grown on 5x5 mm$^2$ (100) MgO and (100) $(LaAlO_3)_{0.3}(Sr_2TaAlO_6)_{0.7}$ (LSAT) substrates using room temperature RF sputtering. The sputtering cathode has the same composition of the deposited films with a 99.9% purity (Testbourne LTD) and dimension of 15x5 cm$^2$. The discharge power density has been varied in the range of 1.5-3.1 W/cm$^2$ with the typical RF frequency of 13.6 MHz. The Ar pressure has been fixed at 4.0 mTorr with a target-substrate distance of 5 cm and a deposition time of 20 minutes. Deposition rates vary from 16 nm/min to 38 nm/min going from 1.5 to 3.1 W/cm$^2$, thus the thicknesses are 320, 515 and 750nm for 130, 170 and 230 W, respectively. The deposition process was then followed by an in-air annealing process consisting in 15°C/min ramp up to the plateau temperature that was maintained for 4 hours. The plateau temperature has been varied in a range from 800°C to 1200°C with steps of 100°C.

## *Characterization of samples*

X-Ray diffraction (XRD) has been performed on annealed samples in order to investigate on the crystal structure of the thin films. The measurements were performed using a D2-Phaser diffractometer (Bruker Corporation, Billerica, Massachusetts, USA) with a Cu−Kα radiation source characterized by a wavelength λ of 1.541 Å. Acquisition time was set to 0.4 s and the obtained spectra have an angular resolution of 0.01°.

In order to investigate the elemental composition of both as-grown and annealed samples, Energy dispersive X-rays spectroscopy (EDX) has been performed on the manufactured thin films. The measurements have been carried out with LaB6 scanning electron microscope (SEM) LEO EVO 50 equipped with X-Rays Energy Dispersion (EDX) system (INCA ENERGY, Oxford Instruments).

Rutherford backscattering spectrometry (RBS) has been performed on as-grown samples. Measurements have been carried out at the Centre of Applied Physics, Dating and Diagnostics (CEDAD) at the Department of Mathematics and Physics "Ennio de Giorgi", University of Salento[24]. An α-particle ($^4$He) beam with an energy of 2MeV has been used, setting the following parameters incident angle 0° and scattering angle 170°.

Temperature-dependent resistance measurements have been performed making use of a Van der Pauw 4-contact geometry. The sample was heated using a hot-plate in the range from room temperature to ≈ 125°C.

# Results and discussion

The NNCO thin films have been deposited on (100) MgO and (100) $(LaAlO_3)_{0.3}(Sr_2TaAlO_6)_{0.7}$ (LSAT) substrates using room temperature RF sputtering with sputtering power density varying in the range 1.5-3.1 W/cm$^2$. After the room temperature sputtering process, the samples have been in-air annealed using the temperature profile shown in the inset to Figure 1, where the plateaux temperature $T_p$ of 900 °C, has been determined analysing the XRD spectra obtained on a series of samples deposited on LSAT substrates and annealed at different $T_p$ values.

## *XRD measurements*

After the annealing, the samples have been analysed by XRD measurements in order to verify the formation of the desired NNCO phase. In Figure 1, the XRD spectra obtained for the samples deposited on MgO substrates at different sputtering power densities are shown. The peaks labelled with an asterisk are related to the MgO substrate, the green diamonds labelled peaks are due to the NNCO phase and those marked with red stars are associated to undesired phases (NNO, $Ni_2CuO_4$, $Nd_2O_3$) [25–28].

Ref. 29 shows that NNCO powders synthesized by the Pechini method, with Cu substitution between 0.15 up to 0.3 on the Ni site, exhibit exclusively the tetragonal NNCO phase, while for higher Cu concentration there is coexistence between the tetragonal NNCO phase and the T' phase of $Ni_2CuO_4$ (T'-NCO)In contrast, the thin films prepared in this work consistently show the coexistence of NNCO and T'-NCO phases, despite the nominal Cu concentration being 0.2 (Figure 1). In particular, for the sample synthesized at 230 W, we can observe many diffraction peaks associated to the tetragonal phase of NNCO, confirming the successful incorporation of Cu in the crystal structure. The presence of T'-NCO as well as other impurities is likely originating to the synthesis route via RF sputtering and post-annealing, for example because of substrate-induced effects or imperfect transfer of the target stoichiometry to the film.

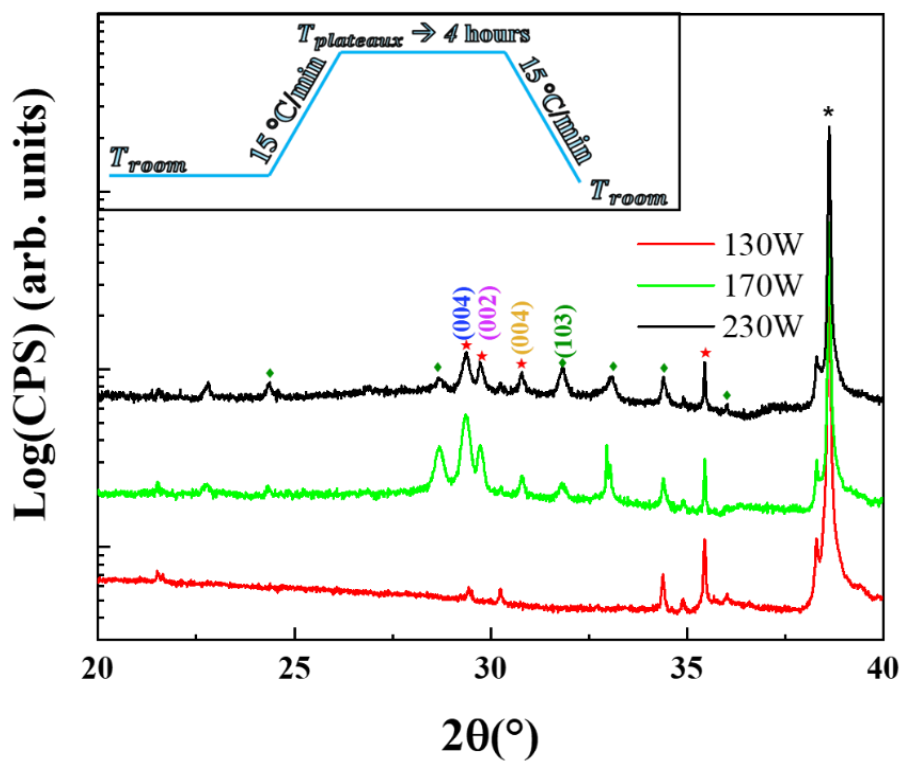

Figure 1 - XRD measurements for samples deposited with different sputtering power discharges: 130 W (red line), 170 W (green line), 230 W (black line). All the spectra are vertically shifted for clarity. In the inset in the left corner the annealing ramp is reported. The asterisks refer the MgO substrate, the green indexed peaks and diamonds refer to the NNCO phase and the red indexed peaks and stars refer to the spurious phases; numbers in brackets refer to NCO (blue Miller indexes), Neodymium oxide (violet Miller indexes), NNO (yellow Miller indexes) and NNCO (green Miller indexes).

## Compositional analysis

To better analyse the influence of the sputtering processes on the elemental composition of the produced samples, we have carried out RBS measurements on the as-grown samples deposited at different sputtering powers on MgO substrates and EDX measurements after and before annealing process.

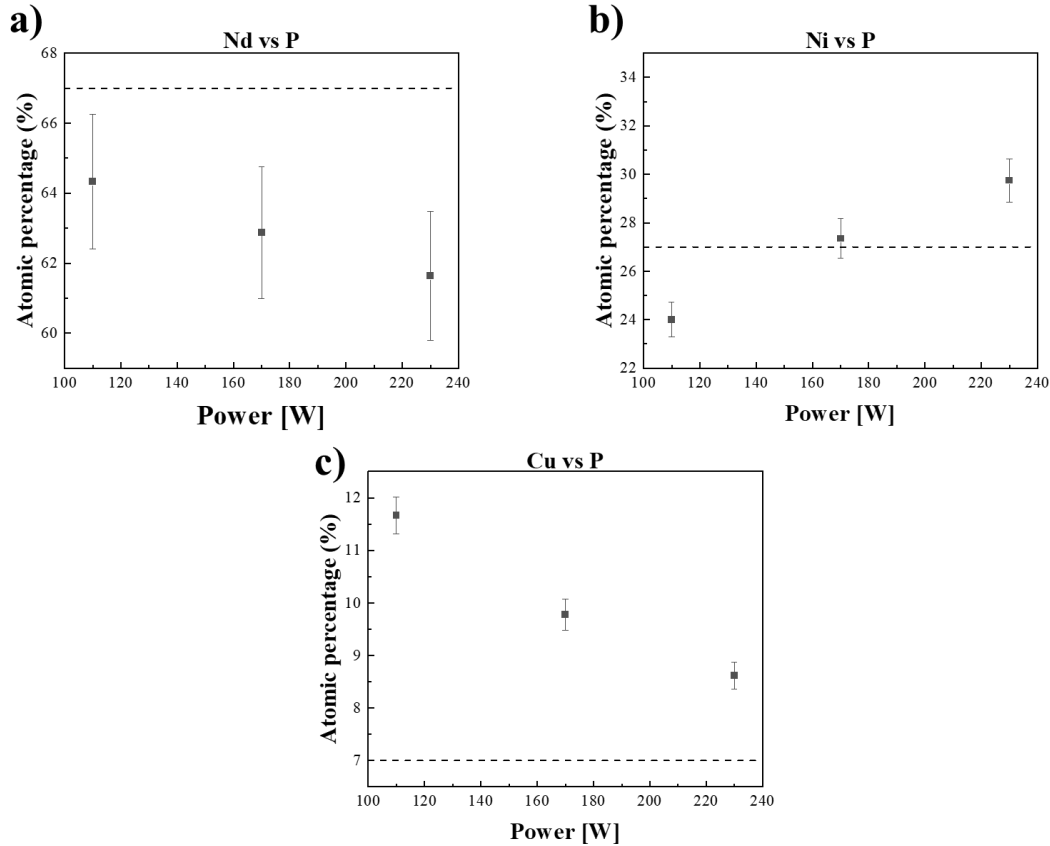

Figure 2 - RBS atomic percentages for concentrations of Nd (a), Ni (b) and Cu (c) as a function of the sputtering power. All the data points are characterized by an error bar evaluated as the 3% of the data. Dashed black line indicates the ratio expected value.

In Figure 2a, 2b and 2c, we show the RBS elemental composition of the films deposited at different sputtering powers for the Nd, Ni and Cu respectively. From Figures 2, for Nd and Ni the observed RBS trends with the power density are opposite. In the case of Nd, the data show a slight tendence to decrease with increasing power densities, although a constant line could be drawn among all the data points inside the reported experimental errors. On the other hand, for Ni, the relative concentration values unambiguously increase with the power density. Finally, for Cu it is possible to observe a reduction of the relative concentration by increasing the power density during the deposition process. The concentration values observed for the Nd, are always below the expected value of 67%, going from about 64% at 1.5 W/cm$^2$ to 61.5% at 3.1 W/cm$^2$, while the Ni percentage goes from 24% to about 30% in the same sputtering power range (expected value of 27%). Always in the same power density range, the Cu percentage goes from about 12% to 9% (expected value of 7%). The large difference between the target and the sample stoichiometry observed for Cu, is not surprising by considering that the sputtering yield of copper is the highest among the heavy elements present in the target. It is interesting to note that this difference decreases with increasing power densities, reaching values close to those in the target at a power density of 3.1 W/cm$^2$. In Figure 3a, we report the calculated RBS Nd/(Ni+Cu) ratios as a function of the power. The values present a decreasing trend with increasing powers, going from 1.8 to 1.6. The value of 1.8 is very close to the expected one (2), apparently indicating the range of lower sputtering powers as the optimal for the obtainment of the desired NNCO phase. On the other hand, in Figure 3b, the reported values of the ratio between the Ni and Cu elemental percentage, clearly show that the samples deposited at lower power are far from the optimal elemental composition. In fact, the Ni/Cu observed ratio at 1.5 W/cm$^2$ is half of that expected (2 instead of 4). The RBS Ni/Cu ratios increase with increasing powers and reach the value of 3.5 at 3.1 W/cm$^2$. Following the previous suggestion obtained by the XRD measurements, this final observation seems to indicate that the critical RBS parameter to focus for the obtainment of the desired NNCO phase is the Ni/Cu ratio instead of the Nd/(Ni+Cu) one. We point out that all the RBS measurements have been performed on as-grown samples.

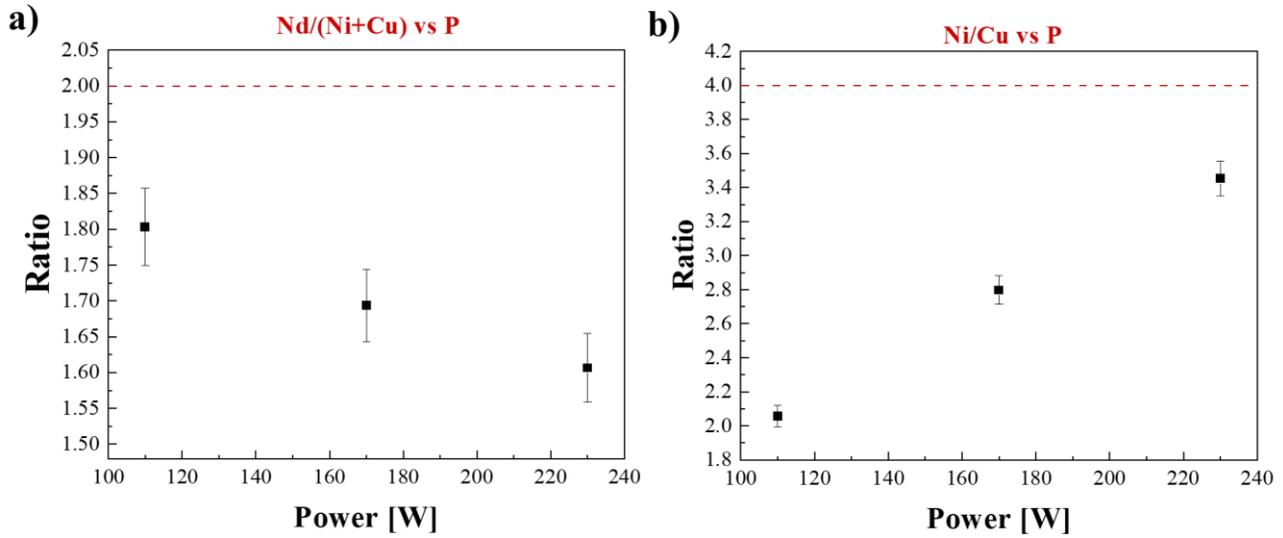

Figure 3 - Atomic percentage ratio evaluated from RBS measurements; in a) Nd/(Ni+Cu) ratio vs. power and in b) Ni/Cu ratio vs. power. We evaluated the error as the 3% of the experimental point value. Dashed red line indicates the ratio expected value.

To confirm that the annealing process has no effect on the sample stoichiometry, we have carried out EDX analysis on films deposited on MgO, before and after the in-air annealing process. In Figures 4a, 4b and 4c, we show these EDX results for the elemental composition of the investigated samples as a function of the discharge power. A larger error is associated to the EDX results compared to RBS, so that some difference between EDX and RBS obtained composition is observed for the thinner samples (i.e. those grown with lower power density) while, for the sample grown at the higher power, the values are essentially the same. Ultimately, these results validate the reliability of the EDX technique for our purpose. Then, moving on to the EDX data, in the case of Nd and Ni, inside the associated experimental errors, all the composition values do not change going from the as-grown to the annealed samples. For Cu, only the sample deposited at very low power density (i.e. 1.5 W/cm$^2$) presents appreciable compositional difference across the annealing process. Again, this observation can be traced back to the difference in the sputtering yields of the materials present in the sputtering target, which is more important at low power values and lose its impact at high power density.

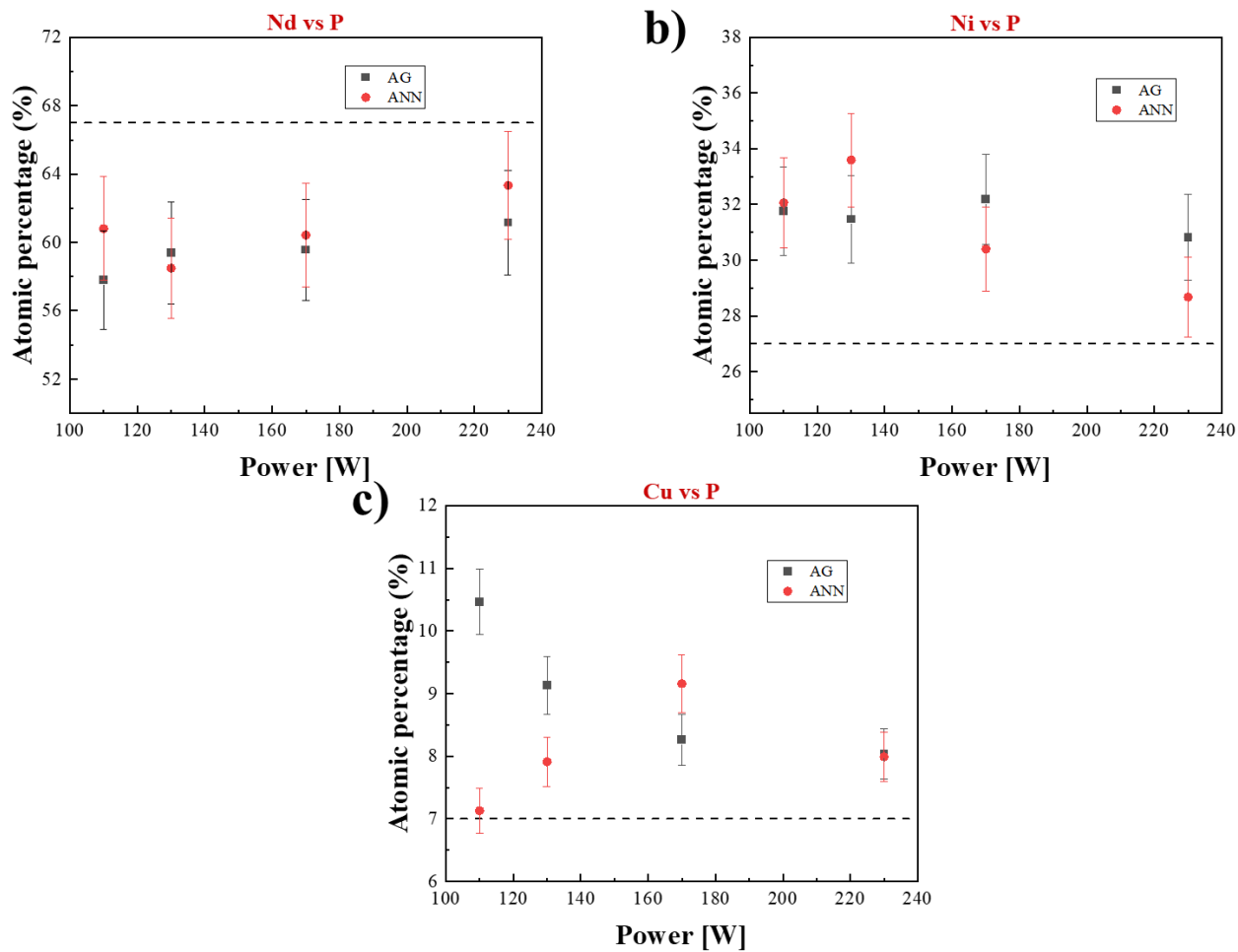

Figure 4 - EDX atomic percentages for concentrations of Nd (a), Ni (b) and Cu (c) as a function of the sputtering power for as grown (AG, grey squares) and annealed (ANN, red dots) samples. We assign an error bar to the measurement equal to 5% of the atomic percentages value for Nd and Ni. For Cu the assigned error bar is equal to 10% of the atomic percentages value. The higher associated error for Cu is justified by the poor amount of this element into the sample which leads to an increase in experimental error.

The XRD, RBS and EDX analysis performed at different power densities, suggest the use of higher power values to obtain the predominant presence of the desired NNCO phase. To confirm this indication, we have finally carried out measurements of the electrical resistivity $\rho$ as a function of the temperature T. The observed $\rho(T)$ curves are shown in Figure 5. The $\rho(T)$ values measured for the sample deposited at 130 W are more than two orders of magnitude higher than those obtained for the samples grown at 170 and 230 W. Moreover, the $\rho$ value at 120 °C for the sample deposited at 230 W is 0.058 Ω*cm and compares well with typical $\rho$ values of 0.015 Ω*cm, observed in bulk NNCO at 300 °C [14]. Data in Figure 5 confirm the previous trend indicated by the XRD, RBS and EDX analyses, showing a substantial reduction of the electrical resistivity in samples deposited at higher powers which is to be associated to the predominant presence of the desired NNCO phase.

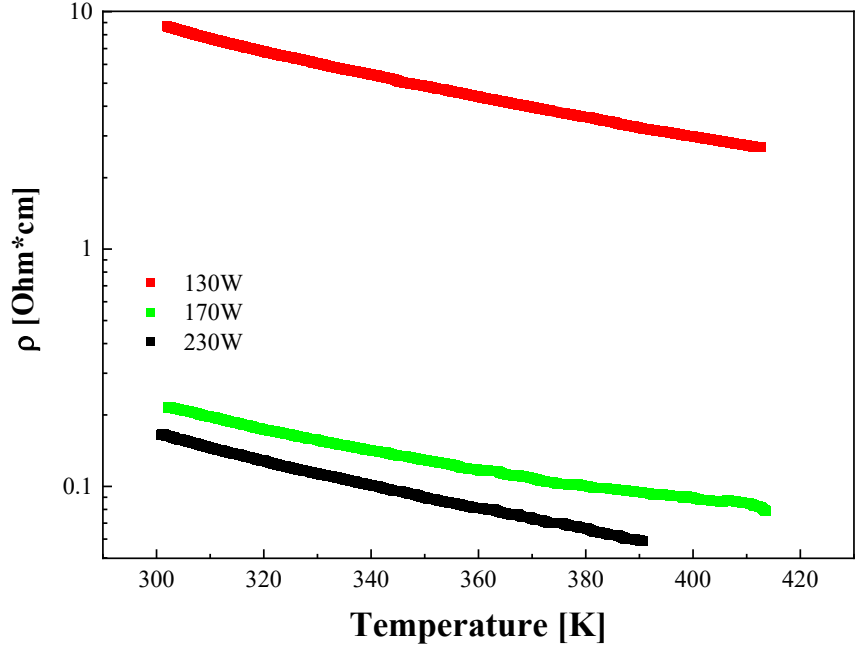

Figure 5 - ρ(T) for NNCO samples grown at 130W (red squares), 170W (green squares) and 230W (black squares)

In layered La based Nickelates, Faucheux et al.[30] observed electric transport properties well described in terms of a Variable Range Hopping (VRH) mechanism. We have analysed the ρ(T) curves of the sample deposited at 230 W in Figure 5, using three possible models:

The first model is the VRH model, in which the temperature dependence of the resistivity is described by the formula

$$\rho(T) = \rho_0 \exp\left(\frac{A}{T^{\frac{1}{4}}}\right) \qquad (i)$$

where $\rho_0$ is the resistivity at high temperature, T is the temperature in K and A is a fitting parameter proportional to $g(E_F)^{-1/4}$ ($g(E_F)$ stands for the density of states at the Fermi level).

The second model is the polaronic activation model [30] where we have

$$\rho(T) = \rho_0 T \exp\left(\frac{E_P}{k_B T}\right) \qquad (ii)$$

Here, $E_P$ is the polaron hopping energy, while $K_B$ is the Boltzman constant.

The third model correspond to the thermal activation model typical of a semiconducting material with

$$\rho(T) = \rho_0 \exp\left(\frac{E_g}{k_B T}\right) \qquad (iii)$$

with $E_g$ equal to the thermal activation energy.

By fitting the data with these three formulas we have obtained the determination coefficients $R^2$ [31] reported in Table 2. Although the $R^2$ values are all very close to 1, the best agreement with the data in Figure 5 is obtained in the case of the VRH model. In agreement with Faucheux et al.[30], this is another indication of the presence of the desired NNCO phase, in the sample deposited at 230 W. As a result of all the analyses (XRD, RBS, EDX, ρ(T)) performed on the samples deposited at different sputtering powers, we have, therefore, a clear indication that the use of high sputtering powers in the adopted deposition process helps to obtain the desired NNCO phase.

Table 1 - Determination coefficients $R^2$ obtained by fitting the $\rho(T)$ curve of the sample deposited at 230 W with the models proposed in the text.

|  | 230W |
|---|---|
| Thermal activation | 0.99908 |
| Polaronic activation | 0.99912 |
| VRH | 0.99955 |

# Conclusions

Using a room temperature sputtering process followed by in air annealing, we have deposited $Nd_2Ni_{0.8}Cu_{0.2}O_{4+\delta}$ thin films in view of their use as cathodes in solid oxide cells. To simplify the deposition process, we adopted a single sputtering oxide target with stoichiometric ratio of the metal ions. This has implied the presence of unwanted phases in the produced samples. We have investigated the influence of the sputtering power density on the relative phase percentages observing, with increasing powers, a reduction of the presence of the spurious phases. The samples have been analysed using XRD, RBS, EDX and $\rho(T)$ measurements. For the films deposited at 230 W (i.e. 3.1 W/cm$^2$), we have observed a predominance of the XRD peaks associated to the NNCO phase with respect to other peaks coming from spurious phases, RBS and EDX values of the elementals concentration close to those expected for $Nd_2Ni_{0.8}Cu_{0.2}O_{4+\delta}$ and resistivity values in agreement with those observed in bulk NNCO. All the performed investigations, therefore, point toward the possibility to obtain, with high sputtering density (around 3.1 W/cm$^2$), thin films with the predominant presence of the NNCO phase with x=0.2, opening the way to future investigations regarding the use of PVD techniques to fabricate innovative solid oxide cells, and, in particular, about their possible scale up in large scale production process. More insights will be provided by Electrochemical Impedance Spectroscopy measurements that will be subject of a future work.

# Acknowledgements

The authors acknowledge financial support under the National Recovery and Resilience Plan (PNRR), Mission 4, Component 2, Investment 1.1, Call for tender No. 1409 published on 14.9.2022 by the Italian Ministry of University and Research (MUR), funded by the European Union – NextGenerationEU– Project Title A-LENS Atomic-level understanding of solid-oxide fuel cell processes– CUP D53D23019420001- Grant Assignment Decree No. 1381 adopted on 01/09/2023 by the Italian Ministry of Ministry of University and Research (MUR).

Part of the results presented in this work were obtained within the framework of the CREA-SUD project ("Advanced and sustainable low and high temperature REversible Cells: unified research and Development of innovative concepts, materials, and Designs"), Grant Agreement No. RSH2A_000030. This project was funded by the Italian Ministry of Ecological Transition (MiTE) under the National Recovery and Resilience Plan (PNRR), Mission 2 "Green Revolution and Ecological Transition", Component 2 "Renewable Energy, Hydrogen, Grid and Sustainable Mobility", Investment 3.5 "Hydrogen Research and Development", funded by the European Union – Next Generation EU (Public Notice No. 0000004 of 23.03.2022)."

The authors also acknowledge the CERIC-ERIC Consortium for the access to the CEDAD laboratory's (Department of Mathematics and Physics "Ennio de Giorgi", University of Salento) experimental facilities and financial support that allowed the RBS measurements presented in this paper.

# Data Availability Statement

The data that support the findings of this study are available from the corresponding author, [N.C.], upon reasonable request.

## *References*

(1) Singh, M.; Zappa, D.; Comini, E. Solid Oxide Fuel Cell: Decade of Progress, Future Perspectives and Challenges. *Int. J. Hydrogen Energy* **2021**, *46* (54), 27643–27674. https://doi.org/10.1016/j.ijhydene.2021.06.020.

(2) Gao, Y.; Zhang, M.; Fu, M.; Hu, W.; Tong, H.; Tao, Z. A Comprehensive Review of Recent Progresses in Cathode Materials for Proton-Conducting SOFCs. *Energy Rev.* **2023**, *2* (3), 100038. https://doi.org/10.1016/j.enrev.2023.100038.

(3) Stambouli, A. B.; Traversa, E. Solid Oxide Fuel Cells (SOFCs): A Review of an Environmentally Clean and Efficient Source of Energy. *Renew. Sustain. Energy Rev.* **2002**, *6* (5), 433–455. https://doi.org/10.1016/S1364-0321(02)00014-X.

(4) Chasta, G.; Himanshu; Dhaka, M. S. A Review on Materials, Advantages, and Challenges in Thin Film Based Solid Oxide Fuel Cells. *Int. J. Energy Res.* **2022**, *46* (11), 14627–14658. https://doi.org/10.1002/er.8238.

(5) Shao, Z.; Zhou, W.; Zhu, Z. Advanced Synthesis of Materials for Intermediate-Temperature Solid Oxide Fuel Cells. *Prog. Mater. Sci.* **2012**, *57* (4), 804–874. https://doi.org/10.1016/j.pmatsci.2011.08.002.

(6) Solovyev, A.; Shipilova, A.; Smolyanskiy, E.; Rabotkin, S.; Semenov, V. The Properties of Intermediate-Temperature Solid Oxide Fuel Cells with Thin Film Gadolinium-Doped Ceria Electrolyte. *Membranes (Basel).* **2022**, *12* (9), 1–7. https://doi.org/10.3390/membranes12090896.

(7) Coppola, N.; Polverino, P.; Carapella, G.; Sacco, C.; Galdi, A.; Montinaro, D.; Maritato, L.; Pianese, C. Optimization of the Electrical Performances in Solid Oxide Fuel Cells with Room Temperature Sputter Deposited Gd0.1ce0.9o1.95 Buffer Layers by Controlling Their Granularity via the in-Air Annealing Step. *Int. J. Hydrogen Energy* **2020**, *45* (23), 12997–13008. https://doi.org/10.1016/j.ijhydene.2020.02.187.

(8) Benamira, M.; Ringuedé, A.; Cassir, M.; Horwat, D.; Lenormand, P.; Ansart, F.; Bassat, J. M.; Viricelle, J. P. Enhancing Oxygen Reduction Reaction of YSZ/La2NiO4+δ Using an Ultrathin La2NiO4+δ Interfacial Layer. *J. Alloys Compd.* **2018**, *746*, 413–420. https://doi.org/10.1016/j.jallcom.2018.02.339.

(9) Tahir, N. N. M.; Baharuddin, N. A.; Samat, A. A.; Osman, N.; Somalu, M. R. A Review on Cathode Materials for Conventional and Proton-Conducting Solid Oxide Fuel Cells. *J. Alloys Compd.* **2022**, *894*, 162458. https://doi.org/10.1016/j.jallcom.2021.162458.

(10) Geisler, H.; Kromp, A.; Weber, A.; Ivers-Tiffée, E. Performance of MIEC Cathodes in SOFC Stacks Evaluated by Means of FEM Modeling. *ECS Meet. Abstr.* **2014**, *MA2014-01* (16), 707–707. https://doi.org/10.1149/ma2014-01/16/707.

(11) Gędziorowski, B.; Cichy, K.; Niemczyk, A.; Olszewska, A.; Zhang, Z.; Kopeć, S.; Zheng, K.; Marzec, M.; Gajewska, M.; Du, Z.; Zhao, H.; Świerczek, K. Ruddlesden-Popper-Type Nd2-XNi1-YCuyO4±δ Layered Oxides as Candidate Materials for MIEC-Type Ceramic Membranes. *J. Eur. Ceram. Soc.* **2020**, *40* (12), 4056–4066. https://doi.org/10.1016/j.jeurceramsoc.2020.04.054.

(12) Zhang, L.; Yao, F.; Meng, J.; Zhang, W.; Wang, H.; Liu, X.; Meng, J.; Zhang, H. Oxygen Migration

(12) and Proton Diffusivity in Transition-Metal (Mn, Fe, Co, and Cu) Doped Ruddlesden-Popper Oxides. *J. Mater. Chem. A* **2019**, *7* (31), 18558–18567. https://doi.org/10.1039/c9ta05893a.

(13) Filonova, E. A.; Pikalova, E. Y.; Maksimchuk, T. Y.; Vylkov, A. I.; Pikalov, S. M.; Maignan, A. Crystal Structure and Functional Properties of Nd1.6Ca0.4Ni1-YCuyO4+δ as Prospective Cathode Materials for Intermediate Temperature Solid Oxide Fuel Cells. *Int. J. Hydrogen Energy* **2021**, *46* (32), 17037–17050. https://doi.org/10.1016/j.ijhydene.2020.10.243.

(14) Lee, K. J.; Choe, Y. J.; Hwang, H. J. Properties of Copper Doped Neodymium Nickelate Oxide as Cathode Material for Solid Oxide Fuel Cells. *Arch. Metall. Mater.* **2016**, *61* (2A), 625–628. https://doi.org/10.1515/amm-2016-0106.

(15) Banner, J.; Akter, A.; Wang, R.; Pietras, J.; Sulekar, S.; Marina, O. A.; Gopalan, S. Rare Earth Nickelate Electrodes Containing Heavily Doped Ceria for Reversible Solid Oxide Fuel Cells. *J. Power Sources* **2021**, *507* (July), 230248. https://doi.org/10.1016/j.jpowsour.2021.230248.

(16) Morales, M.; Pesce, A.; Slodczyk, A.; Torrell, M.; Piccardo, P.; Montinaro, D.; Tarancón, A.; Morata, A. Enhanced Performance of Gadolinia-Doped Ceria Diffusion Barrier Layers Fabricated by Pulsed Laser Deposition for Large-Area Solid Oxide Fuel Cells. *ACS Appl. Energy Mater.* **2018**, *1* (5), 1955–1964. https://doi.org/10.1021/acsaem.8b00039.

(17) Coppola, N.; Sami, H.; Rehman, U.; Carapella, G.; Polverino, P.; Montinaro, D.; Martinelli, F.; Granata, V.; Galdi, A.; Maritato, L.; Pianese, C.; Studi, D.; Sa, F. Large Area Solid Oxide Fuel Cells with Room Temperature Sputtered Barrier Layers : Role of the Layer Thickness and Uniformity in the Enhancement of the Electrochemical Performances and Durability. *Int. J. Hydrogen Energy* **2023**, No. 48(7). https://doi.org/10.1016/j.ijhydene.2023.04.170.

(18) Yunhui Gong, Rajankumar L. Patel, Xinhua Liang, Diego Palacio, Xueyan Song, John B. Goodenough, and K. H. Atomic Layer Deposition Functionalized Composite SOFC Cathode La0.6Sr0.4Fe0.8Co0.2O3-δ -Gd0.2Ce0.8O1.9: Enhanced Long-Term Stability. *Chem. Mater.* **2013**, *25*, 4224–4231.

(19) Coppola, N.; Polverino, P.; Carapella, G.; Ciancio, R.; Rajak, P.; Montinaro, D.; Martinelli, F.; Maritato, L.; Pianese, C. Large Area Deposition by Radio Frequency Sputtering of Gd0.1Ce0.9O1.95 Buffer Layers in Solid Oxide Fuel Cells: Structural, Morphological and Electrochemical Investigation. *Materials (Basel).* **2021**, *14*, 5826. https://doi.org/https://doi.org/10.3390/ma14195826.

(20) Coppola, N.; Ur Rehman, S.; Carapella, G.; Braglia, L.; Vaiano, V.; Montinaro, D.; Granata, V.; Chaluvadi, S. K.; Orgiani, P.; Torelli, P.; Maritato, L.; Aruta, C.; Galdi, A. Effects of In-Air Post Deposition Annealing Process on the Oxygen Vacancy Content in Sputtered GDC Thin Films Probed via Operando XAS and Raman Spectroscopy. *ACS Appl. Electron. Mater.* **2024**. https://doi.org/10.1021/acsaelm.4c00992.

(21) Coppola, N.; Polverino, P.; Carapella, G.; Sacco, C.; Galdi, A.; Ubaldini, A.; Vaiano, V.; Montinaro, D.; Maritato, L.; Pianese, C. Structural and Electrical Characterization of Sputter-Deposited Gd0.1Ce0.9O2−δ Thin Buffer Layers at the Y-Stabilized Zirconia Electrolyte Interface for IT-Solid Oxide Cells. *Catalysts* **2018**, *8* (12). https://doi.org/10.3390/catal8120571.

(22) Siddall, G. Handbook of Thin Film Technology. *Thin Solid Films.* 1971, p 473. https://doi.org/10.1016/0040-6090(71)90064-2.

(23) Guarino, A.; Patimo, G.; Vecchione, A.; Di Luccio, T.; Nigro, A. Fabrication of Superconducting Nd2-XCexCuO 4±δ Films by Automated Dc Sputtering Technique. *Phys. C Supercond. its Appl.* **2013**, *495*, 146–152. https://doi.org/10.1016/j.physc.2013.09.010.

(24) Calcagnile, L.; Quarta, G.; Elia, M. D.; Rizzo, A. A New Accelerator Mass Spectrometry Facility in Lecce , Italy. *NIMB* **2004**, *223–224*, 16–20. https://doi.org/10.1016/j.nimb.2004.04.007.

(25) Tarutin, A. P.; Lyagaeva, Y. G.; Vylkov, A. I.; Gorshkov, M. Y.; Vdovin, G. K.; Medvedev, D. A. Performance of Pr2(Ni,Cu)O4+δ Electrodes in Protonic Ceramic Electrochemical Cells with Unseparated and Separated Gas Spaces. *J. Mater. Sci. Technol.* **2021**, *93*, 157–168. https://doi.org/10.1016/j.jmst.2021.03.056.

(26) Munawar, T.; Sardar, S.; Nadeem, M. S.; Mukhtar, F.; Manzoor, S.; Ashiq, M. N.; Khan, S. A.; Koc, M.; Iqbal, F. Rational Design and Electrochemical Validation of Reduced Graphene Oxide (RGO) Supported CeO2-Nd2O3/RGO Ternary Nanocomposite as an Efficient Material for Supercapacitor Electrodes. *J. Appl. Electrochem.* **2023**, *53* (9), 1853–1868. https://doi.org/10.1007/s10800-023-01885-0.

(27) Soliman, T. S. Investigation of Structural and Optical Features of Polyvinyl Alcohol Films Doped with Nd2O3 Nanoparticles for UV Shielding. *Opt. Quantum Electron.* **2024**, *56* (7). https://doi.org/10.1007/s11082-024-07112-0.

(28) Funke, M.; Blum, M.; Glaum, R.; Elbali, B. Darstellung, Kristallstruktur Und Spektroskopische Charakterisierung von NiP 4O 11 Und CaNiP 2O 7. *Zeitschrift fur Anorg. und Allg. Chemie* **2004**, *630* (7), 1040–1047. https://doi.org/10.1002/zaac.200400077.

(29) Wang, C.; Soga, H.; Okiba, T.; Niwa, E.; Hashimoto, T. Construction of Structural Phase Diagram of Nd 2 Ni 1- x Cu x O 4 + δ and Effect of Crystal Structure and Phase Transition on Electrical Conduction Behavior ☆. *Mater. Res. Bull.* **2019**, *111* (August 2018), 61–69. https://doi.org/10.1016/j.materresbull.2018.10.036.

(30) Faucheux, V.; Audier, M.; Pignard, S. Physical Properties of Epitaxial La 2 NiO 4 + d Thin Films. *Appl. Surf. Sci.* **2006**, *252*, 5504–5507. https://doi.org/10.1016/j.apsusc.2005.12.136.

(31) Draper, N. R.; Smith, H. *Applied Regression Analysis, 3rd Edition*; Wiley Series in Probability and Statistics, 1998.